\documentclass{natureMOD}
\usepackage{amsmath,amssymb, amstext}
\usepackage[T1]{fontenc}
\usepackage[latin1]{inputenc}
\usepackage{graphicx}
\usepackage{amssymb}
\usepackage{dcolumn}
\usepackage{color}
\usepackage{bm}
\usepackage{siunitx}
\usepackage[normalem]{ulem}
\usepackage{url}
\usepackage[ colorlinks = true,
            linkcolor = red,
            urlcolor  = red,
            citecolor = blue,
            anchorcolor = red]{hyperref}
\usepackage[colorinlistoftodos]{todonotes}
\usepackage{bibentry}

\begin{document}

%% give the tite here
\title{Interplay of filling fraction and coherence in symmetry broken graphene p-n junction}

%% give all the author name here
\author{Arup Kumar Paul$^1$$^*$, Manas Ranjan Sahu$^1$$^*$, Chandan Kumar$^1$, Kenji Watanabe$^{2}$, Takashi Taniguchi$^{2}$ and Anindya Das$^1$\footnote{anindya@iisc.ac.in}}

\maketitle

\begin{affiliations}
\item Department of Physics,Indian Institute of Science, Bangalore, 560012, India.
\item National Institute for Materials Science, Namiki 1-1, Ibaraki 305-0044, Japan.
\end{affiliations}


\begin{abstract}
The coherence of quantum Hall (QH) edges play the deciding factor in demonstrating an electron interferometer, which has potential to realize a topological qubit. A Graphene $p-n$ junction (PNJ) with co-propagating spin and valley polarized QH edges is a promising platform for studying an electron interferometer. However, though a few experiments have been attempted for such PNJ via conductance measurements, the edge dynamics (coherent or incoherent) of QH edges at a PNJ, where either spin or valley symmetry or both are broken, remain unexplored. In this work, we have carried out the measurements of conductance together with shot noise, an ideal tool to unravel the dynamics, at low temperature ($\sim$ 10mK) in a dual graphite gated hexagonal boron nitride ($hBN$) encapsulated high mobility graphene device. The conductance data show that the symmetry broken QH edges at the PNJ follow spin selective equilibration. The shot noise results as a function of both $p$ and $n$ side filling factors ($\nu$) reveal the unique dependence of the scattering mechanism with filling factors. Remarkably, the scattering is found to be fully tunable from incoherent to coherent regime with the increasing number of QH edges at the PNJ, shedding crucial insights into graphene based electron interferometer.
\end{abstract}

\maketitle

\section{Introduction:}
Ever since the realization that the charge and energy are carried by the edge states in a QH system, the interest of edge dynamics have surged both theoretically and experimentally. The understanding of edge dynamics makes an electron interferometer suitable for exploring exotic phenomena like fractional statistics, quantum entanglement and non-abelian excitations~\cite{neder2006unexpected,neder2007interference,law2006electronic,feldman2006detecting,stern2010non}. A Graphene $p-n$ junction %in the quantum Hall (QH) regime 
naturally harboring co-propagating electron and hole like edge states offers an ideal platform~\cite{zhang2005experimental,neto2009electronic,abanin2006spin,ozyilmaz2007electronic,williams2007quantum,rickhaus2015snake,taychatanapat2015conductance,klimov2015edge, handschin2017giant,nakaharai2011gate} to study the edge or equilibration dynamics. The equilibration of such edge states is predicted to be facilitated by inter-channel tunnelling via either incoherent or coherent scattering mechanism~\cite{tworzydlo2007valley,abanin2007quantized,li2008disorder,long2008disorder,low2009ballistic,chen2011dephasing,frassdorf2016graphene,lagasse2016theory,ma2018graphene}  depending on the microscopic details of the interface. As suggested by Abanin et. al.~\cite{abanin2007quantized}, for a graphene PNJ interface with random disorders, the edge mixing is expected to be dominated by the incoherent process. In the opposite limit, a cleaner PNJ interface %where strong inter-valley scattering at the PNJ plays a crucial role in the mixing process
~\cite{tworzydlo2007valley,morikawa2015edge,frassdorf2016graphene,wei2017mach}
is supposed to exhibit coherent scattering. A cleaner PNJ interface is also very intriguing for studying the equilibration dynamics as it has spin and valley symmetry broken polarized QH edges~\cite{zhang2006landau,jiang2007quantum,young2012spin,young2014tunable}. Though there are several conductance measurements~\cite{zimmermann2017tunable,amet2014selective,wei2017mach} showing spin-selective partial equilibration of the edges, but the equilibration dynamics for symmetry broken QH edges at a PNJ is still unknown. 

Shot noise is a quintessential tool to unravel the equilibration dynamics of a junction and it is usually characterized by Fano factor ($F$), which is the ratio of the actual noise to its Poissonian counterpart. For coherent and incoherent scattering, $F = (1-t)$ and $t(1-t)$, respectively~\cite{buttiker1992scattering,texier2000effect,blanter2000shot,abanin2007quantized,oberholzer2001shot}, with $t$ being the average transmission of the PNJ. So far, shot noise studies~\cite{kumada2015shot,matsuo2015edge} at graphene PNJ in the QH regime have been performed on $Si/SiO_2$ substrate-based devices, where the spin-valley symmetry broken conductance plateaus are not observed %due to limited quality of the devices. 
and the measured Fano~\cite{kumada2015shot,matsuo2015edge} fairly agrees with the incoherent model~\cite{abanin2007quantized} due to disorder limited interface. Besides, the shot noise measurements are focused around the lowest filling factor ($\nu = \pm2$) and hence the dependence of $F$ on filling factors ($\nu$) is lacking. More importantly, there are no shot noise studies for spin-valley symmetry broken QH edges at graphene PNJ.  

With this motivation, we have carried out the conductance together with shot noise measurements at a PNJ realized in a dual graphite gated $hBN$ encapsulated high mobility graphene device. From the conductance measurement, we show that the spin and valley degeneracies of the edge states are completely lifted and at the PNJ the edge states undergo spin selective partial equilibration. Our shot noise data %at $\sim 10mK$ 
as a function of filling factors shows the following important results: (1) The Fano strongly depends on the filling factors. It monotonically increases with $p$ side filling factors, whereas it slowly varies with $n$ side filling factors. (2) For lower values of $p$ side filling factors ($\nu_p \leq 2$), the variation of Fano matches well with the calculated Fano based on incoherent scattering model, whereas for higher values of $p$ side filling factors ($\nu_p \geq 4$) Fano follows the coherent scattering model. These results reveal a crossover of scattering process from incoherent to coherent regime in the equilibration of %polarized 
QH edges, which has not been observed in the previous shot noise studies~\cite{kumada2015shot,matsuo2015edge}.

\begin{figure*}[ht!]
\includegraphics{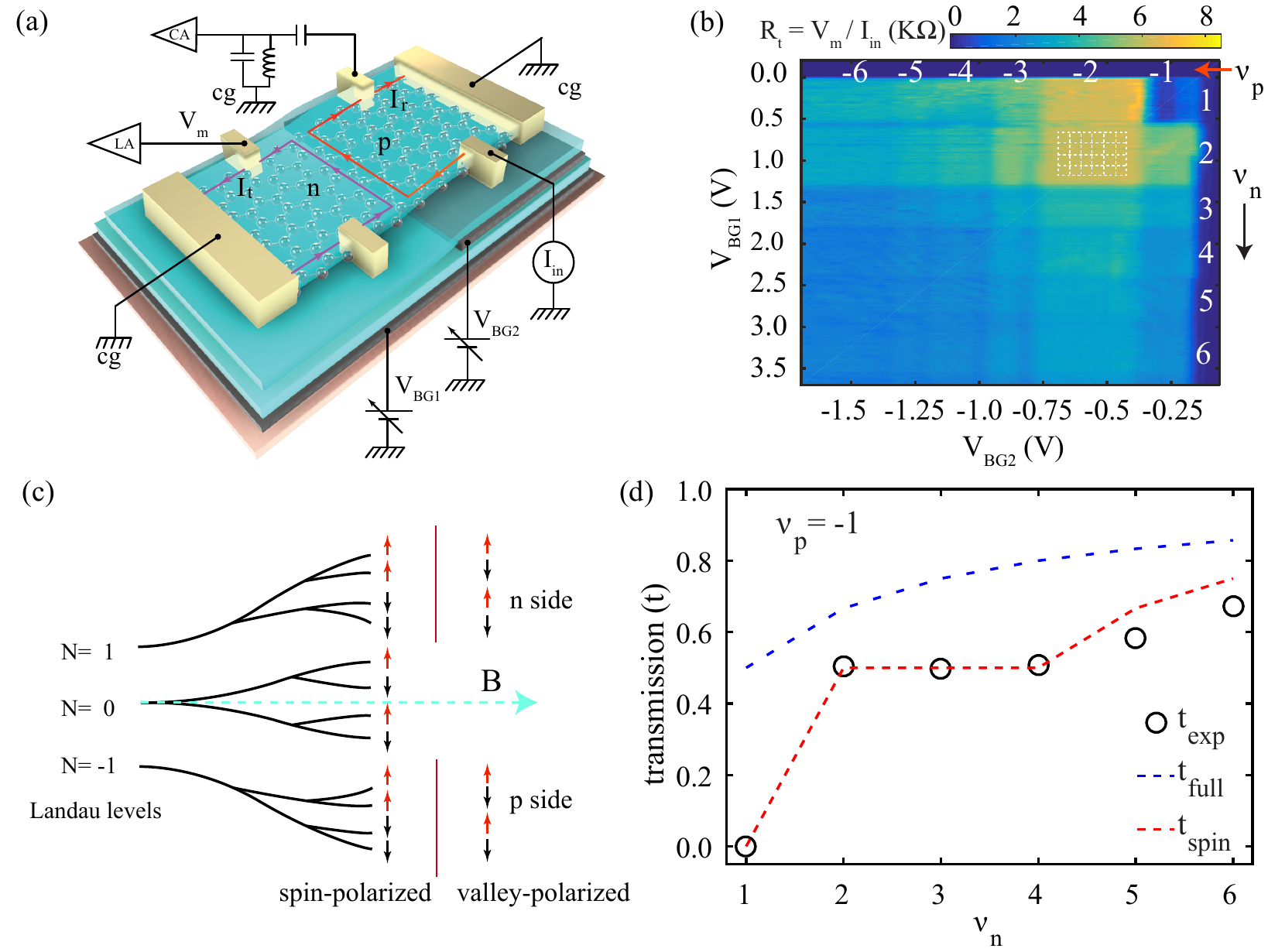}%
\caption{\label{fig:epsart} \textbf{Measurement set-up and junction transmittance:} (a) Schematics of the device and measurement setup: The encapsulated graphene flake is positioned on top of two bottom graphite gates $BG1$ and $BG2$, which are connected to the gate voltages $V_{BG1}$ and $V_{BG2}$.  The chirality of the edge states during measurement, for $p$ and $n$ doped region, is shown by the red and purple arrowed lines. For both conductance as well as shot noise measurements  excitation current $I_{in}$ is injected at the $p$ side. The transmitted current ($I_t$) at the $n$ side is determined by measuring the voltage $V_m$ with a lock-in amplifier (LA). The shot noise generated at the PNJ is measured at the $p$ side using a resonant tank circuit followed by a cryogenic amplifier (CA). The extreme left and right contacts were grounded to the dilution mixing chamber plate serving as cold ground (cg). (b) The trans-resistance $R_t=V_m/I_{in}$, as a function of $V_{BG1}$ and $V_{BG2}$. $R_t$ shows a checkerboard-like pattern corresponding to the different combinations of $p$ and $n$ side filling factors $\nu_p$ and $\nu_n$, shown in the white vertical and horizontal axis, respectively. The white dashed region on ($\nu_p,\nu_n$) = (-2,2) plateau, are to show how noise data were taken on several different points of the same plateau. (c) Spin configuration of the edge states %at both sides of the junction, 
for two different ways of LL degeneracy lifting: spin and valley polarized ground states. (d) Measured transmission ($t$) (open circles) of the PNJ as a function of filling factor $\nu_n$ for $\nu_p=-1$. The calculated $t$ for full equilibration and partial equilibration with spin splitting configuration is shown by blue and red dashed lines, respectively. %As can be seen the experimental data matches very well with the selective equilibration model.
}
\end{figure*}

\section{Results:}
\subsection{Measurement set-up:}
The schematics of our device with the measurement setup is shown in Figure 1(a). The PNJ device is fabricated by placing an hBN encapsulated graphene on top of two graphite gates $BG1$ and $BG2$, each of which can independently control the carrier density of one half of the graphene (details in supporting information (SI) figure SI-1). The PNJ (width $\sim$ 10 $\mu$m) is obtained at the interface of $BG1$ and $BG2$ by applying opposite voltages to the gates. During our entire measurement, the $BG1$ ($BG2$) side is maintained as $n$ ($p$) doped, by setting gate voltage $V_{BG1}>0$ ($V_{BG2}<0$). When a perpendicular magnetic field is applied to the graphene, chirally opposite QH edge states co-propagating along the PNJ, are created as shown by the colored arrow lines in Fig. 1(a). As shown in the figure, the current ($I_{in}$) injected at the $p$ doped region is carried by clockwise edge-states towards the PNJ. After partitioning at the PNJ, the transmitted current ($I_{t}$) at the $n$ doped region and reflected current ($I_r$) at the $p$ doped region is carried by the outgoing anti-clockwise and clockwise edge-states, respectively. The shot noise generated due to partitioning at the PNJ is carried by both the transmitted and reflected paths. %both $I_t$ and $I_r$. 
To measure $I_t$ and the shot noise, the measurement setup consists of two parts: 1) A low frequency ($\sim 13$ Hz) part, which determines $I_t$ by measuring the voltage drop $V_m$ at $n$ doped region, with a Lock-in amplifier (LA) as shown in Fig. 1(a) (also see SI-2(a)). 2) A high frequency shot noise measurement part, where a DC current ($I_{in}$) is injected at $p$ doped region and the generated noise is measured at reflected side using LCR resonant circuit at $\sim 765$ kHz as shown in Fig. 1(a) (described in detail in SI-2(b)). All the measurements were performed at $8$ T magnetic field inside a cryo-free dilution fridge (with base temperature $\sim 10$ mK), whose mixing chamber plate serves as the cold ground (cg in Fig.1(a)). 

\subsection{Conductance measurement:}
Figure 1(b) shows trans-resistance, $R_{t} = {V_m}/{I_{in}}$ as a function of $V_{BG1}$ and $V_{BG2}$. The plot shows plateau-like features creating a checkerboard pattern for different combinations of $\nu_p$ and $\nu_n$, where $\nu_p$ and $\nu_n$ are the $p$ and $n$ side filling factors, respectively (details in SI-3(a)). The transmittance $t = I_{t}/I_{in}$ of each plateau is determined from the $R_t$ as $t = |\nu_n|R_{t}/\frac{h}{e^2}$, where $V_m = I_tR_h$ and $R_ h$ $ = \frac{h}{e^2} / |\nu_n|$ is the QH resistance of the $n$ doped region. Figure 1(d) shows the measured values of $t$ (open circles) as function $\nu_n$ for $\nu_p =-1$ with the corresponding theoretical values considering full equilibration\cite{abanin2007quantized}; $t={\nu_n}/(\nu_p + \nu_n)$ (blue dashed line), and spin selective partial equilibration\cite{amet2014selective,zimmermann2017tunable,kumar2018equilibration}; $t=\frac{1}{\nu_p}[\frac{\nu_{p\uparrow}\nu_{n\uparrow}}{\nu_{p\uparrow}+\nu_{n\uparrow}} + \frac{\nu_{p\downarrow}\nu_{n\downarrow}}{\nu_{p\downarrow}+\nu_{n\downarrow}}]$ (red dashed line), where $\nu_{p\uparrow}$($\nu_{p\downarrow}$) and $\nu_{n\uparrow}$($\nu_{n\downarrow}$) are the total numbers of up (down) spin edge channels of the $p$ and $n$ doped region, respectively (SI-5). For spin selective equilibration, two possible sequences of spin polarization of the edge states (valley or spin polarized  ground state) are shown in Fig. 1(c)\cite{amet2014selective}. The red dashed line in Fig. 1(d) is based on the spin structure for the spin-polarized ground state and it is in very good agreement with the measured $t$. Note that the other spin sequence also gives good agreement with the experimental data. For simplicity, we will be presenting only one of them (spin-polarized  ground state) throughout the manuscript. The measured $t$ and the calculated values based on spin selective equilibration for other plateaus are also in very good agreement and are shown in SI-5. %within $\leq10\%$ mismatch as shown in SI}. 
%Thus, the conductance measurement establishes that the symmetry broken PNJ follows the spin selective partial eqilibration.

\subsection{Shot noise measurement:}
In this section, we present the results of our shot noise measurement. The  shot noise generated at the PNJ is measured at the reflected side ($p$ side) as a function of $I_{in}$, as shown in Fig. 1(a). In general, measured current noise ($S_I$) consists of both thermal and shot noise and follows the expression:
\begin{equation}
S_I=2eI_{in}F^{*}[coth(\frac{eV_{sd}}{2k_BT})-\frac{2k_BT}{eV_{sd}}]
\end{equation}
where $V_{sd}$ is the applied bias voltage across the PNJ, $T$ is the temperature and $k_B$ is Boltzmann constant. For $eV_{sd}>k_BT$ shot noise dominates over thermal noise and $S_I$ becomes linear with $I_{in}$ as shown in Figure 2(a) for ($\nu_p,\nu_n$) = $(-2,2)$, $(-3,3)$ and $(-4,4)$ filling factor plateaus. %(visible in Fig. 1(b). 
The red lines in Fig. 2(a) are the fit using equation (1). The slopes of the fit have been used to determine the normalized noise magnitude ($F^{*} = \frac{S_I}{2eI_{in}}$). For obtaining Fano ($F = \frac{S_I}{2eI_t}$) we just follow $F  = F^{*}/t$, which is conventionally used to characterize the noise and used in the previous shot noise studies on graphene PNJ\cite{matsuo2015edge,kumada2015shot}. Figure 2(b) shows the histogram of $F$ obtained from the noise data taken at several points ($\sim$ 50) on each checkerboard (plateau) as shown by the white dotted squares in Fig. 1(b) for ($\nu_p,\nu_n$) = $(-2,2)$. The histograms are fitted with the Gaussian function as shown by the solid red lines in Fig. 2(b) for ($\nu_p,\nu_n$) = $(-2,2)$, $(-3,3)$ and $(-4,4)$ plateaus. It can be seen that the histograms have a maximum at a certain value of $F$ (mean value), which depends on the filling factors $(\nu_p,\nu_n)$. The noise data and the corresponding histograms for some other plateaus are shown in SI (SI-14, SI-15 and SI-16). We should note that to pinpoint the exact scattering mechanism, the accuracy of the extracted Fano is very essential. This accuracy depends on the amplifier gain, noise from the contacts as well as on enough statistics. In SI-4, the precise gain calibration and in SI-13, the measured contact noise as a function filling factors are shown. The contact noise has been subtracted in the histogram plots shown in Fig. 2(b) as well as in SI.

\begin{figure*}[ht]
\includegraphics[width=1\textwidth]{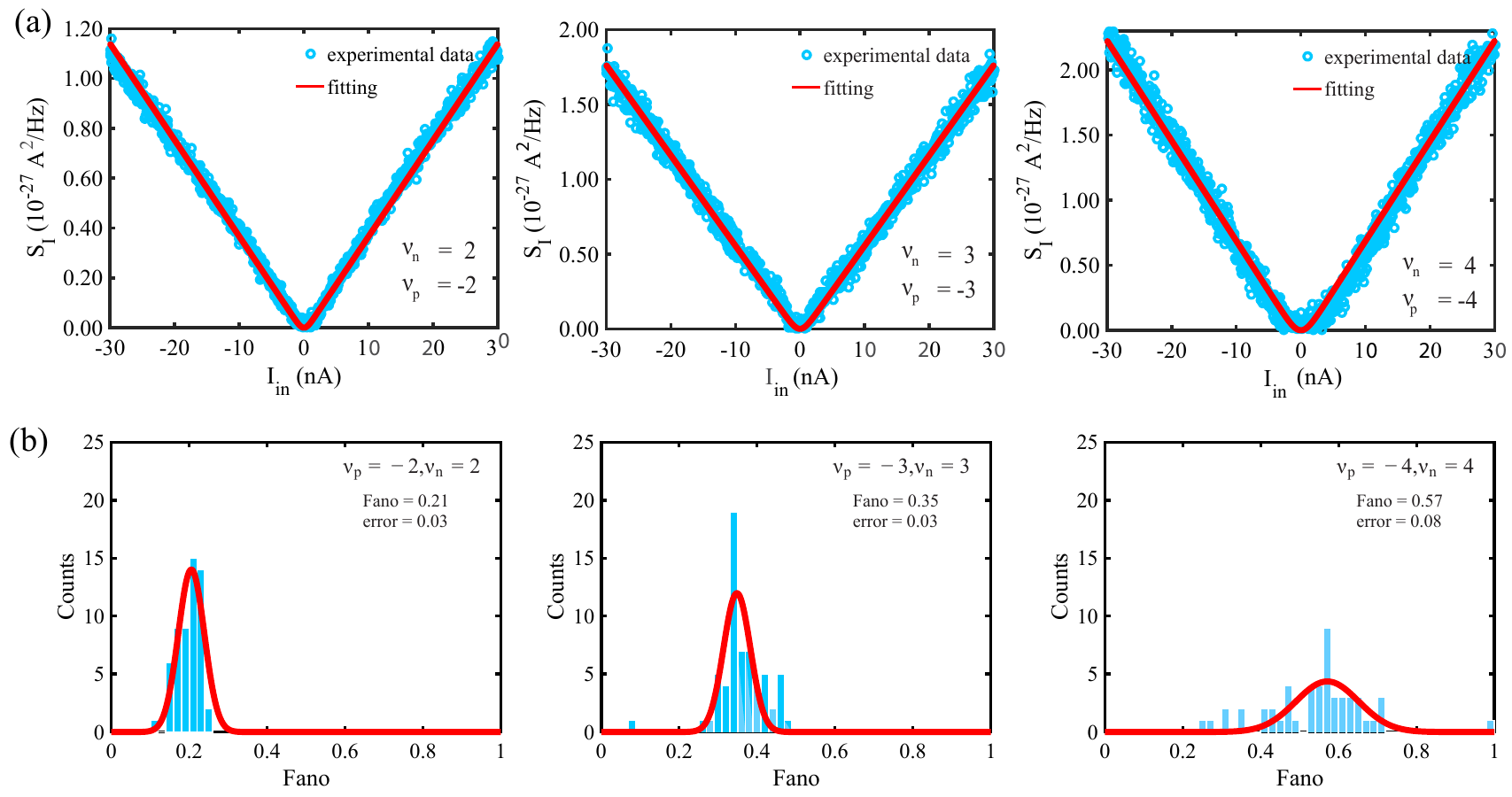}% Here is how to import EPS art
\caption{\label{fig:epsart}\textbf{Shot noise and extraction of Fano factor :} (a) Measured current noise ($S_I$ ) generated by the PNJ as a function of injected current $I_{in}$, for ($\nu_p,\nu_n$) = (2,-2), (3,-3) and (4,-4)
plateaus. The solid red lines are the fit with Eqn. 1 to extract the Fano factor. %($F = S_{I}/2eI_{t}$)
(b) Histogram of Fano factor for ($\nu_p,\nu_n$) = (2,-2), (3,-3) and (4,-4) plateaus taken at several points ($\sim$ 50) as shown in Fig. 1(b). The solid lines are the Gaussian fit to extract the mean value of $F$ and its standard deviation.}
\end{figure*}

\section{Discussion:}
\subsection{Fano versus filling factor:}
The  measured values of $F$ (mean value) as a function of filling factor are shown in Fig. 3 as open circles with the error bars (standard deviations of Gaussian fit in Fig. 2b). In Fig. 3(a) and 3(b), $F$ is plotted as a function of $\nu_p$ while the $n$ side filling factor is kept fixed at $\nu_n=2$ and $\nu_n=5$, respectively. It can be seen that $F$ increases monotonically from $\sim$ 0.05 to 0.6 with increasing $\nu_p$. Similarly, in Fig. 3(c) and 3(d), $F$ is plotted as a function of $\nu_n$ while the $p$ side filling factor is kept fixed at $\nu_p=-2$ and $\nu_p=-4$, respectively. However, in this case, the $F$ does not increase monotonically with $\nu_n$, rather slowly varies around $\sim$ 0.2 and 0.6 for $\nu_p=-2$ and $\nu_p=-4$, respectively. Similar dependence of $F$ on $\nu_p$ or $\nu_n$ for other fixed values of $\nu_n$ or $\nu_p$ is shown in SI (SI-11 and SI-12).

\subsection{Comparison with theoretical models:}
To understand the above results, we theoretically calculate $F$ for coherent and incoherent processes. In coherent scattering, the injected hot carriers from $p$ side (Fig. 1(a)) coherently scatter to the $n$ side and the inter-channel scattering can be described by scattering matrix approach\cite{texier2000effect,buttiker1992scattering}. In this  case, $F$ follows as $(1-t)$ similar to that of a quantum point contact (QPC) (details in SI-6). Furthermore, for our symmetry broken PNJ, we also impose the constraints that the two opposite spin channels do not interact with each other~\cite{amet2014selective,kumar2018equilibration}. Thus the Fano can be written as 
$ F_{coherent}=(\nu_{p\uparrow}t_{\uparrow}(1-t_{\uparrow})+\nu_{p\downarrow}t_{\downarrow}(1-t_{\downarrow}))/(\nu_{p\uparrow}t_{\uparrow}+\nu_{p\downarrow}t_{\downarrow})
 $, where $t_{\uparrow}=\nu_{n\uparrow}/(\nu_{n\uparrow}+\nu_{p\uparrow}$) and  $t_{\downarrow}=\nu_{n\downarrow}/(\nu_{n\downarrow}+\nu_{p\downarrow})$ are the transmittance of up and down spin channels, respectively (SI-5). The calculated $F_{coherent}$  is shown as red dashed lines in Fig. (3) %for the spin sequence of Fig. 1(c) 
(SI-6). $F_{coherent}$ increases with $\nu_p$ but decreases with $\nu_n$, which can be qualitatively understood as the transmittance of the PNJ decreases and increases with $\nu_p$ and $\nu_n$, respectively. For incoherent scattering we consider both the quasi-elastic and inelastic processes\cite{texier2000effect,abanin2007quantized}. In quasi-elastic case, known as chaotic cavity model, the injected hot carriers from $p$ side %elastically 
scatters to the $n$ side and subsequently scatters back and forth %between the co-propagating edges 
due to the presence of disorders along the PNJ giving rise to double-step distribution~\cite{texier2000effect,abanin2007quantized,kumada2015shot,oberholzer2001shot,le2010energy,altimiras2010non}. Following Abanin et. al\cite{abanin2007quantized} the expression for $F$ is $t(1-t)$ and for our symmetry broken PNJ Fano can be written as $ F_{incoherent}=(\nu_{p\uparrow}t_{\uparrow}^2(1-t_{\uparrow})+\nu_{p\downarrow}t_{\downarrow}^2(1-t_{\downarrow}))/(\nu_{p\uparrow}t_{\uparrow}+\nu_{p\downarrow}t_{\downarrow})
 $. The calculated $F_{incoherent}$ is shown as blue dashed lines in Fig. (3) (SI-6). $F_{incoherent}$ remain almost constant around $F \sim 0.2$ and much smaller in magnitude compared to $F_{coherent}$. Note that the calculated values of Fano using inelastic scattering as described by Abanin et. al.\cite{abanin2007quantized} are very similar in magnitude with the quasi-elastic case (SI-10(e)).

\begin{figure*}[ht]
\includegraphics[width=1\textwidth]{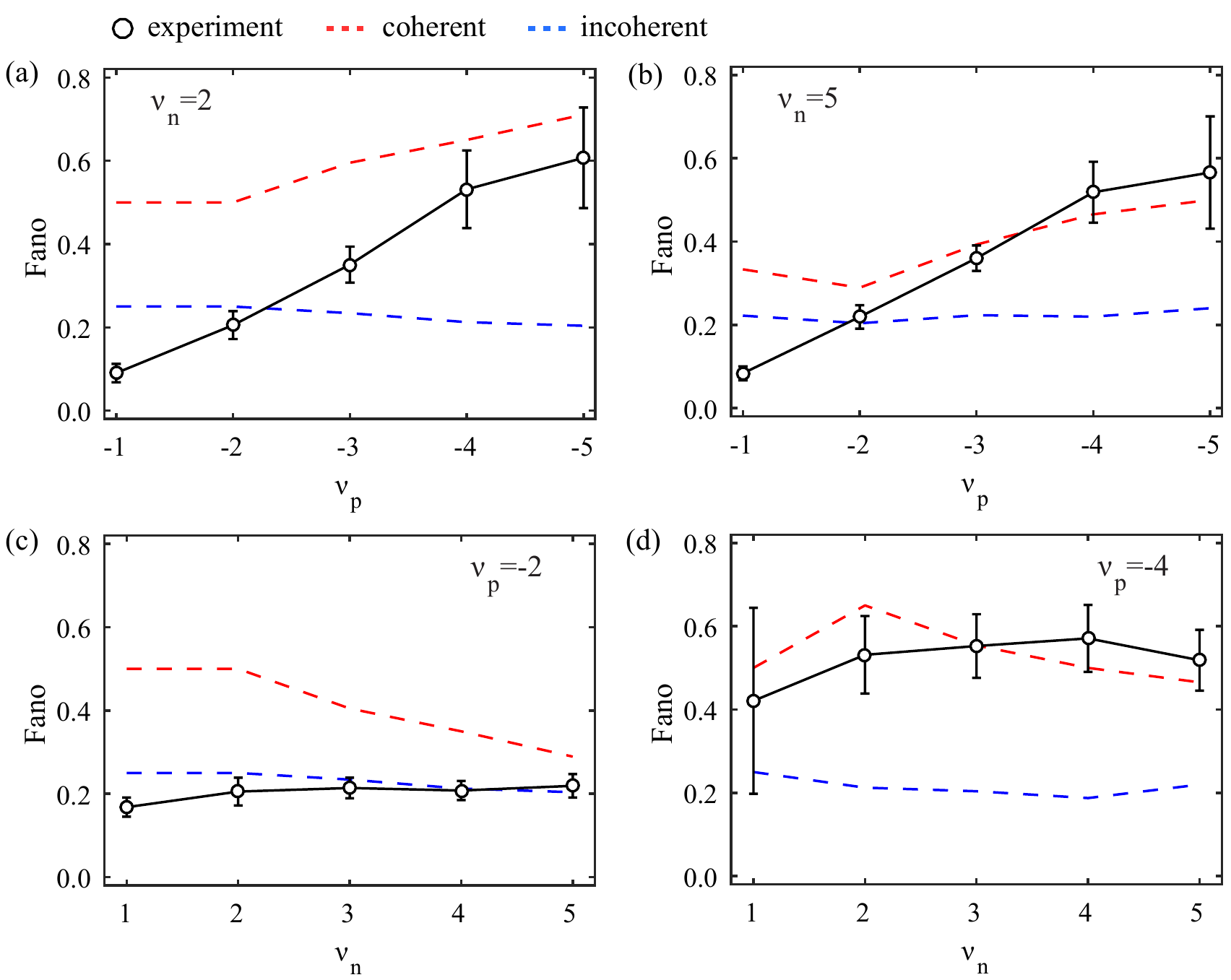}% Here is how to import EPS art
\caption{\label{fig:epsart}\textbf{Fano versus filling factor:} Fano as a function of filling factor $\nu_p$ for $\nu_n=2$ (a) and $\nu_n=5$ (b). Open circles with error bars represent the experimentally measured Fano. Red and blue dashed lines correspond to the calculated Fano for coherent and incoherent scattering (quasi-elastic), respectively. %As can be seen for lower values of $\nu_p (=-1,-2)$ experimental Fano is closer to incoherent model and then monotonically approaches towards the coherent model with increasing $\nu_p$, except for $\nu_p=-6$. 
Fano as a function of filling factor $\nu_n$ for $\nu_p=-2$ (c) and $\nu_n=-4$ (d). Red and blue dashed lines correspond to the calculated Fano for coherent and incoherent scattering (quasi-elastic), respectively.}
\end{figure*}

The monotonic increase of $F$ with $\nu_p$ in Fig. 3(a) and 3(b) is in contradiction with the incoherent scattering model and is consistent with the coherent case except for lower values of $\nu_p$. However, the measured $F$ with $\nu_n$ for $\nu_P=-2$ matches very well with the incoherent scattering, but for $\nu_p=-4$ it perfectly matches the coherent medel. This suggests there is a cross over from incoherent to the coherent regime with the increasing number of edge channels at the $p$ side. This is further verified in Fig. 4, where the measured $F$ plotted as a function of $|\nu_p| = |\nu_n|$ and increases monotonically from $\sim$ 0.2 to 0.55 (open circles), whereas $F_{incoherent}$ (blue dashed line) and $F_{coherent}$ (red dashed line) remain constant around $\sim$ 0.25 and 0.5, respectively. We believe that the screening might be playing a big role in dynamics as observed in GaAs based 2DEG\cite{altimiras2010tuning,le2010energy,gurman2016dephasing,ofek2010role,zhang2009distinct}. The coherent scattering dominates as the screening increases with more number of participating edges at PNJ. %Further theoretical studies are required to quantitatively understand our shot noise results.

\section{Conclusion:} 
In summary, we have carried out conductance together with shot noise measurement on a high-quality graphene $p-n$ junction, for the first time, with spin and valley symmetry broken quantum Hall edges. We have shown that the conductance data follows the spin-selective partial equilibration, and most importantly, our shot noise data reveals the intricate dependence of Fano on filling factors with a crossover in dynamics from incoherent to the coherent regime, which can not be obtained from the conductance measurements. These results will help to design future electron optics experiments using the polarized QH edges of graphene.

\begin{figure*}[t]
\includegraphics{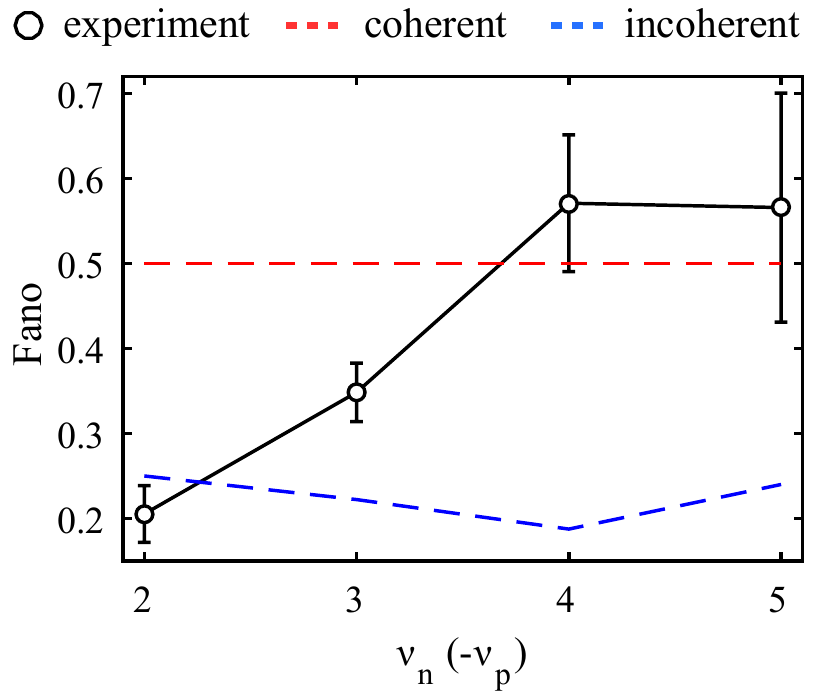}% Here is how to import EPS art
\caption{\label{fig:epsart}\textbf{Crossover from incoherent to coherent scattering regime:} Fano as function of filling factors when $|\nu_n|=|-\nu_p|$. Red and blue dashed lines corresponds to the calculated Fano for coherent and incoherent scattering (quasi-elastic), respectively. The cross over from incoherent to coherent regime is seen with increasing filling factors.}
\end{figure*}

\section{Methods:}
\subsection{Device fabrication:} To make the encapsulated device, the $hBN$ and graphene, as well as the graphite flakes for bottom gates, were exfoliated from bulk crystals on  $Si/SiO_2$ substrates. Natural graphite crystals were used for exfoliating graphene and the graphite flakes.  The suitable flakes for the device were first identified under an optical microscope and then sequentially assembled with the residue-free polycarbonate-PDMS stamp technique~\cite{purdie2018cleaning,pizzocchero2016hot,zomer2014fast}. We have used $15 nm$ and $25$ nm thick hBN flakes for encapsulating the graphene flake and $10\sim15 nm$ thick graphite flakes for the bottom gates. To make the metal edge contacts on the device, first, the contacts were defined with e-beam lithography technique. Then along the defined region, only the top hBN flake was etched out using $CHF_3-O_2$ plasma. After that $Cr(2 nm)/Pd(10 nm)/Au(70 nm)$ was deposited using thermal evaporation. 

\subsection{Shot noise set- up:} To measure the shot noise, first, the voltage noise generated from the device is filtered by a superconducting resonant LC tank circuit, with resonance frequency at 765 kHz and bandwidth 30 kHz~\cite{sahu2019enhanced,srivastav2019universal}. The filtered signal is then further amplified by the HEMT cryo amplifier followed by a room temperature amplifier. The amplified signal is then fed to a spectrum analyzer, to measures the r.m.s of the signal. The gain of the amplifier chain is determined from the temperature dependence of the thermal noise of $\nu_p=-2$ filling factor plateau, while the $n-side$ is in the insulating state. The thermal noise measurement is carried out using the same noise circuit.

\section{Acknowledgement:}
This is a pre-print of an article published in \textbf{Communications Physics}. The final authenticated version is available online at:   https://doi.org/10.1038/s42005-020-00434-x.  Authors thank Dr. Sumilan Banerjee and Dr. Tanmoy Das for the valuable comments. AD thanks DST (DSTO-2051) and acknowledges the Swarnajayanti Fellowship of the DST/SJF/PSA-03/2018-19 for the financial support. KW and TT acknowledge support from the Elemental Strategy Initiative conducted by the MEXT, Japan and and the CREST (JPMJCR15F3), JST.

\section{Contribution :}
AP and MRS contributed to the device fabrication, data acquisition, and
analysis. CK contributed in the noise setup and preliminary experiments. AD contributed in conceiving the idea and designing the experiment, data interpretation, and analysis. KW and TT synthesized the hBN single crystals. All authors contributed in writing the manuscript. 

\pagebreak

\bibliography{ref}

\end{document}